\documentclass{edp-conf}
\usepackage{graphicx}
\begin{document}
\def\la{\buildrel<\over\sim}
\def\ga{\buildrel>\over\sim}
\def\bbe{\rm B$\rightleftharpoons$Be\ }
\def\bbes{\rm B$\rightleftharpoons$Be$\rightleftharpoons$Be-shell\ }

\TitreGlobal{SF2A 2005}

\title{THE LONG-TERM H$\alpha$ EMISSION LINE VARIATION IN $\alpha$~Eri}

\author{Vinicius, M.M.F.}\address{Instituto de Astronomia, Geof\'\i sica e 
Ci\^encias Atmosf\'ericas da Universidade de S\~ao Paulo, Brazil}
\author{Leister, N.V.$^1$}
\author{Zorec, J.}\address{Institut d'Astrophysique de Paris, UMR7095 CNRS, 
Universit\'e Pierre \& Marie Curie}
\author{Levenhagen R.S.$^1$}

\runningtitle{$\alpha$~Eri: Long-term emission variation}

\index{Vinicius, M.M.F.}
\index{Leister, N.V.}
\index{Zorec, J.}
\index{Levenhagen, R.S.}

\maketitle

\begin{abstract}The long-term variation of the H$\alpha$ line in $\alpha$~Eri
has 14-15 years cyclic \bbe phase transitions. The disc formation time scales, 
interpreted as the periods during which the H$\alpha$ line emission increases 
from zero to its maximum, agree with the viscous decretion model. On the other 
hand, the time required for the disc dissipation ranges from 6 to 12 years 
which questions the viscous disc model predictions.
\end{abstract}

\section{Introduction}

 Among the outstanding characteristics of Be stars are their phase \bbes 
transitions evidenced by the Balmer line emission intensity changes. These 
changes can be ought either to variations of the physical structure and size 
of a more or less permanent circumstellar envelope (CE), or to the 
creation of a new CE during mass ejection events that undergo the central star. 
Some keys on the CE formation can be obtained from the time scales of their 
emission line variation, which can be related directly with the disc formation. 
In this contribution we present the cyclic long-term H$\alpha$ line emission 
changes observed in $\alpha$~Eri.

\section{Cyclic long-term H$\alpha$ line emission changes}

 We collected in the literature the spectroscopic and photometric records of 
the H$\alpha$ line variation and studied them as a function of time. We also
take advantage of new observations of the $\alpha$~Eri H$\alpha$ line emission 
changes, to derive information on time scales that characterize complete \bbe 
transitions. The time scales of these cycles matter to understand the disc 
formation mechanisms (Porter 1999, Okazaki 2001).\par
 If the evolution of the CE in $\alpha$~Eri, i.e. formation and subsequent 
dissipation, has to be understood in terms of a viscous decretion disc model 
(Porter 1999, Okazaki 2001), from the expected average viscous time scales 
$t\sim60/\alpha$ days (Clark et al. 2003) we would derive two different series 
of values of the viscosity coefficient $\alpha$. The time scales implied by 
the CE formation, 2 years roughly, we would obtain $\alpha \sim$ 0.08 which is 
of the same order as expected for CE envelopes studied by Clark et al. (2003). 
The time required by the decrease from maximum to zero is either 10 years in 
the 1974-1989 cycle or 6 years during the 1991-2000 cycle. This imply $0.02 
\la\alpha\la0.03$ which is about 4 times smaller than implied in the formation 
process. We may then wonder whether the bulk redistribution of material in the 
circumstellar disc occurs only as a consequence of a viscous redistribution of 
angular momentum.\par

\section{Conclusion}

 The long-term H$\alpha$ line emission variation in $\alpha$~Eri has a 
long-term, roughly cyclical \bbe\ phase transitions every 14-15 years. The 
disc formation time scale, i.e. period when the H$\alpha$ line emission 
increases from zero to its maximum, agrees with the viscous decretion model. 
On the other hand, the time required for the disc dissipation ranges from 6 to 
12 years, which may question the viscous disc model.\par

\begin{figure}[h]
\centering
\includegraphics[height=6cm,width=12cm]{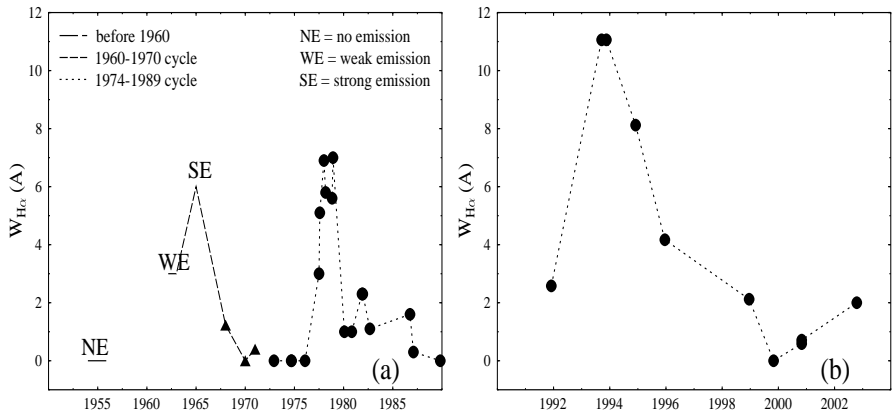}
\caption{Long-term variation of the H$\alpha$\ line emission in $\alpha$~Eri. 
(a): Qualitative and quantitative estimations of the emission strength before
1990 collected in the literature. (b): New equivalent widths of the H$\alpha$ 
line emission obtained after 1990. Different markers were used to identify 
the emission cycles better.}
\end{figure}

\end{document}